\documentclass[9pt,conference]{IEEEtran}

\usepackage{amsmath, graphicx}
\usepackage{subcaption, siunitx, booktabs}
\usepackage{enumitem}
\usepackage{bm, amsfonts}
\usepackage{cite}
\usepackage{amsmath}
\usepackage{multicol, multirow}
\usepackage{comment}

\def\BibTeX{{\rm B\kern-.05em{\sc i\kern-.025em b}\kern-.08em
    T\kern-.1667em\lower.7ex\hbox{E}\kern-.125emX}}

\def\x{{\bm x}}

\def\z{{\bm z}}

\def\D{{\mathcal D}}

\begin{document}
\title{Timbre Difference Capturing in\\ Anomalous Sound Detection}
%
\author{\IEEEauthorblockN{Tomoya Nishida, Harsh Purohit, Kota Dohi, Takashi Endo, Yohei Kawaguchi}
\IEEEauthorblockA{\textit{Research and Development Group, Hitachi, Ltd.}}
}
\maketitle
\begin{abstract}
This paper proposes a framework of explaining anomalous machine sounds in the context of anomalous sound detection~(ASD). 
While ASD has been extensively explored, identifying how anomalous sounds differ from normal sounds is also beneficial for machine condition monitoring.
However, existing sound difference captioning methods require anomalous sounds for training, which is impractical in typical machine condition monitoring settings where such sounds are unavailable.
To solve this issue, we propose a new strategy for explaining anomalous differences that does not require anomalous sounds for training.
Specifically, we introduce a framework that explains differences in predefined timbre attributes instead of using free-form text captions.
Objective metrics of timbre attributes can be computed using timbral models developed through psycho-acoustical research, enabling the estimation of how and what timbre attributes have changed from normal sounds without training machine learning models.
Additionally, to accurately determine timbre differences regardless of variations in normal training data, we developed a method that jointly conducts anomalous sound detection and timbre difference estimation based on a k-nearest neighbors method in an audio embedding space.
Evaluation using the MIMII DG dataset demonstrated the effectiveness of the proposed method.
\end{abstract}

\section{Introduction}
\label{sec:intro}
\vspace{-5pt}
Anomalous sound detection (ASD)~\cite{koizumi2020description} is the task of identifying whether the sound emitted from a target machine is normal or anomalous.
This leads to automatic detection of mechanical failures, which is useful for machine condition monitoring.
Since anomalous sounds are challenging to collect, ASD is often challenged in scenarios where only normal sounds are available for training, known as Unsupervised ASD (UASD)~\cite{kevin2024whydo}.
Various methods and techniques have been developed for UASD in the literature~\cite{koizumi2019unsupervised, suefusa2020anomalous, giri2020unsupervised, dohi2021flow-based, lopez2021ensemble, wilkinghoff2024self, kevin2024whydo, LvAITHU2024}.

(U)ASD only detects the condition (normal or anomalous) of a given sound without specifying how the anomalous sound differs from normal sounds.
Consequently, further analysis is necessary to determine the cause of the anomaly and whether repairs are needed.
Identifying how the anomalous sound differs from normal sounds can make this analysis easier, as it may indicate the type of machine malfunction.
We will refer to such differences observed in anomalous sounds as the ``anomalous difference''.

One such approach is presented in \cite{Takeuchi2023audiodifference}, where a model is trained to output captions describing differences between normal and anomalous sounds. 
A similar task was also explored in ~\cite{Takeuchi2023audiodifference} for general environmental sounds and sound events.
However, these methods do not align with the typical UASD problem setting because: 
1) They require ground truth difference captions and paired data of normal and anomalous sounds for training, 
which is impractical since anomalous samples are unavailable in UASD, and creating captions for paired data is laborious.
2) They only compare a given pair of audio samples, whereas UASD aims to detect whether an anomalous sample deviates from the entire normal data distribution.
Therefore, the anomalous difference explanation should also be based on how the anomalous sample deviates from the normal data distribution.
Comparing just a pair of samples makes it difficult to capture the true deviation in the anomalous sound due to the variability in normal sounds.
3) They solely focus on anomalous difference captioning without performing UASD.
Since anomalous difference captioning is typically conducted after UASD, captions that align with the UASD results are preferred.

\begin{figure}[!t]
\begin{center}
        \includegraphics[width=0.6\hsize,clip]{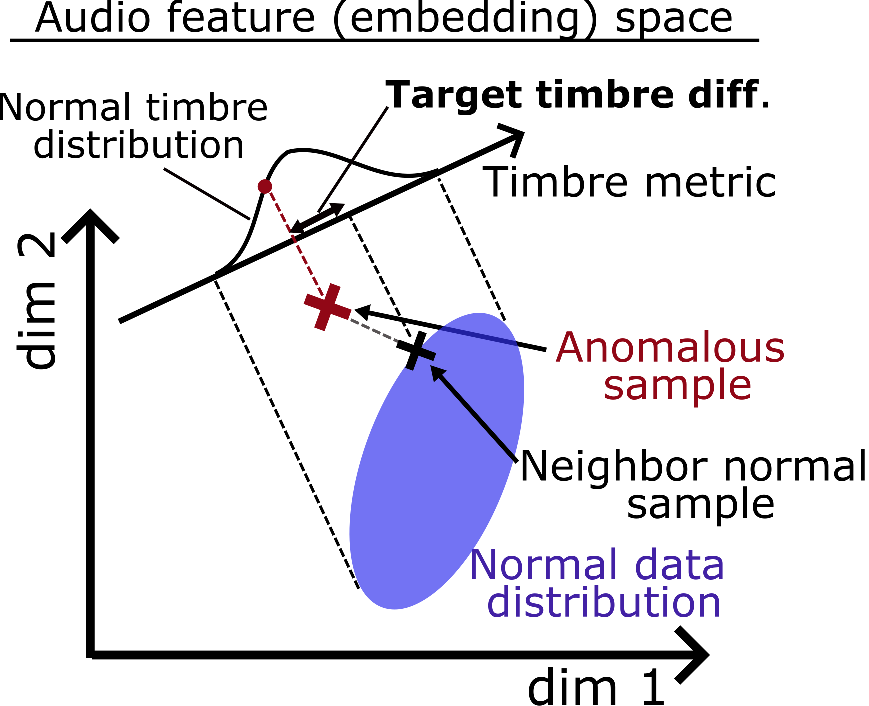}
	\caption{Illustration of data distribution in audio feature (or embedding) space and aim of proposed method. When comparing timbre metric value of anomalous sample to whole normal data, differences cannot be determined (red dot in normal timbre distribution). By comparing timbre metric only with neighbor normal samples in feature space, timbre difference can be determined (Target timbre diff.).}
	\label{fig:method_abs}
        \vspace{-25pt}
\end{center}
\end{figure}
\begin{figure*}[!t]
	\begin{center}
        \includegraphics[width=0.85\hsize,clip]{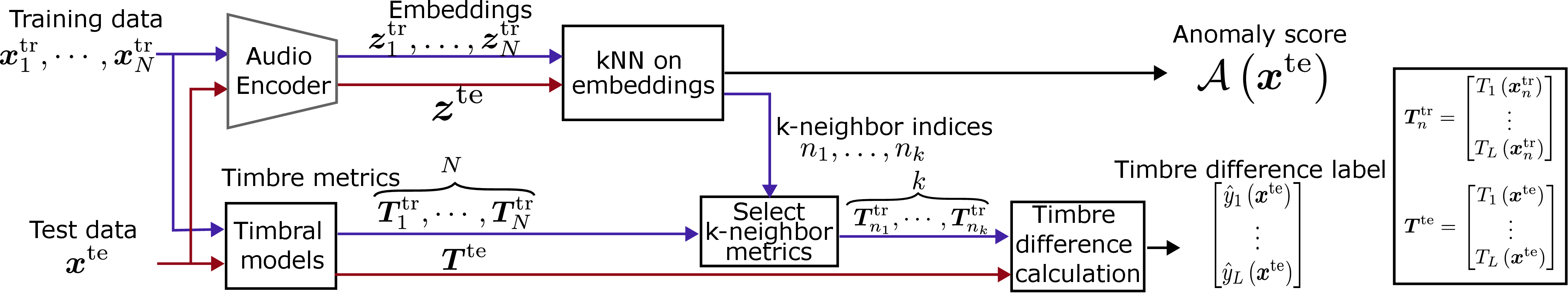}
	\caption{Overview of proposed joint UASD and timbre difference capturing method in inference phase.}
        \vspace{-20pt}
	\label{fig:method}
	\end{center}
\end{figure*}
To address these issues, we propose a new strategy for anomalous difference explanation that suits UASD problem settings and a method for joint UASD and anomalous difference explanation.
\textbf{Specifically, 
we introduce an anomalous difference explanation framework that solely explains differences in predefined timbre attributes such as sharpness or boominess, instead of using free-form text captions (Contribution 1)}.
The goal is to determine whether each timbre attribute has increased, decreased, or remained unchanged in the anomalous sample.
We term this framework "timbre difference capturing``.
Timbre has been factorized into multiple timbral attributes through psycho-acoustical research~\cite{zwicker2013psychoacoustics}, and objective metrics that quantify these attributes have been developed~\cite{jensen2002timbre, pearce2017timbral}.
Using these metrics, we can infer the impression anomalous sounds will give to humans without model training, which could help solve issue 1).
These timbral metrics have also been used to detect machine malfunctioning sounds~\cite{mian2022efficient, Ota2023Anomalous}, 
where \cite{Ota2023Anomalous} listed timbre attributes that are related to machine malfunctions through interviews to factory machine inspectors.
This exemplifies that explaining anomalous differences using timbre attributes is informative for humans.

However, even if timbral metrics can be computed for anomalous sounds during inference, determining which timbre attribute has changed from normal data and how is still challenging due to the variations in normal sounds caused by different machine conditions, recording conditions or environmental noise.
\textbf{To solve this problem, 
we further propose a joint UASD and timbre difference capturing method that can estimate timbre differences despite the variations in normal samples (Contribution 2).}
Here, we hypothesize that the timbre difference caused by anomalies would appear in the difference between the anomalous sound and the most similar normal samples.
Based on this hypothesis, we use the k-nearest neighbors method in an audio embedding space to find normal training samples that are similar to the test samples. 
Then, we perform UASD and timbre difference capturing using the found similar normal training samples.


\section{Joint UASD and timbre difference capturing}
\label{sec:problemsetting}
\vspace{-7pt}
In this section, we define our proposed task of joint UASD and timbre difference capturing.
We assume the typical UASD problem setting for both training and inference phases, as described in previous studies~\cite{koizumi2020description, Kawaguchi2021description, dohi2022description, Dohi2023description}.
For training data, we only have access to normal data, denoted as $\{\x_n^{(\text{tr})}\}_{n=1}^{n=N}$.
Additional information for each normal sample, such as machine speed or microphone location, may or may not be included in the training data.
During inference, both normal and anomalous samples are provided without additional attribute information.
The goal of UASD is to classify each sample as normal or anomalous during inference.

Since anomalous data are unavailable for training, existing sound difference captioning methods~\cite{Tsubaki2023AudioChange, Takeuchi2023audiodifference} are inapplicable because the models cannot be trained.
To solve this, we propose a new framework that focuses solely on describing the changes in predefined timbre attributes instead of using free-form captions. 
Timbre is a multi-dimensional attribute and various descriptive phrases, such as sharpness and boominess, are used to express the impression of a sound~\cite{zwicker2013psychoacoustics}.
Research has modeled these timbral attributes as objective metrics~\cite{jensen2002timbre}, resulting in timbral models that numerically represents the extent to which humans perceive each attributes.
For example, the Audio Commons project~\cite{pearce2017timbral} includes several timbral models widely used in related research. 
Timbre attribute and timbre metrics have been used in machine condition monitoring~\cite{minemura2018acoustic, mian2022efficient, Ota2023Anomalous}, indicating its relevance to machine malfunctions.
Especially, \cite{Ota2023Anomalous} listed timbre attributes related to machine malfunctions through interviews to factory machine inspectors, which exemplifies anomalous difference explanation based on timbre attributes can be informative.

Our proposed task of timbre difference capturing is as follows.
First, we predefine $L$ timbral attributes $l=1,\dots,L$ that are known to be related to machine malfunctions.
In this paper, we select the $5$ attributes considered relevant to machine malfunctions in ~\cite{Ota2023Anomalous}: {\bf 1) Sharpness:} Sharp or shrill sensation, {\bf 2) Roughness:} Buzzing, harsh, raspy sound quality, {\bf 3) Boominess:} Booming sensation, often perceived as a low-pitch vibration, {\bf 4) Brightness:} Bright sensation, and {\bf 5) Depth:} Emphasized low-frequency component.
For further details of these attributes and models, refer \cite{Ota2023Anomalous}.

The goal of timbre difference capturing for an anomalous sample $\x^{(\text{te})}$ is to determine, for each timbre attribute $l=1,\dots ,L$, whether the timbre impression has increased, decreased, or not changed by the anomaly.
For example, if a machine malfunction causes an additional buzzing sound, the Roughness attribute may increase while other attributes remain unchanged.
Identifying these changes can explain how the anomalous sound differs from the normal sound as perceived by humans. 
We denote the estimated timbre difference as $y_l \in \{1,0,-1\}, l=1,\dots,L$, where $1, 0, -1$ stands for increased, no change, and decreased, respectively.
Thus, each test sample will be given $L$ labels with values being one of $\{1, 0, -1\}$, making the task a combination of multi-label classification and ordinal classification.

In this task, timbre metrics can be computed for anomalous samples by timbral models without training.
However, since normal sounds vary, simply calculating the timbral metrics for anomalous sounds does not determine the timbre difference.
This issue is addressed by the proposed method.
Finally, since this task only explains differences in a limited number of timbre attributes, the expressiveness is limited compared to free-form captions.
While solving this task alone is still informative, we leave this limitation to be solved in future work.


\section{Proposed method}
\label{sec:method}
\vspace{-7pt}
\subsection{Overview}
\vspace{-5pt}
By using timbral models, we can compute timbre metrics for both normal training data and anomalous data found during inference.
However, this alone does not directly identify timbre differences due to the variety in normal sounds.
What truly needs to be identified is how the anomalous sound differs from specific normal sounds occurred under identical conditions, except for the normal/anomalous status.
These conditions include the machine's operational mode, recording conditions, or environmental noise.
These factors affect the recorded sound as well as the normal/anomalous status, meaning that sound differences due to such condition variations needs to be ruled out in anomalous difference explanation.
To solve this issue, we hypothesize that normal sounds under identical conditions are likely to be the samples that are the most similar to the anomalous sound.
Thus, we presume that timbre differences can be estimated more accurately by comparing the timbre metrics of the anomalous sound only with these similar normal sounds.
To realize this idea, we use an audio encoder to extract embeddings of each sample and use the nearest neighbor samples in the embedding space as the similar normal sounds.
We illustrate this concept in Fig.~\ref{fig:method_abs}.

Fig.~\ref{fig:method} summarizes the proposed method.
For jointly conducting UASD and timbre difference capturing, we assume a system that first extracts embeddings of given audio samples by an arbitrary audio encoder.
This is followed by a k-nearest neighbor~(knn) detector that identifies $k$ neighbor samples of the test data from $\{\x_n^{(\text{tr})}\}_{n=1}^{n=N}$.
The audio encoder can be a pre-trained model, a model trained from scratch with the normal training data, or a pre-trained model with fine-tuning.
Since the preparation and training of audio encoders are arbitrary, we omit explanations for this part.
In the following subsections, we explain how UASD and timbre difference capturing are conducted in inference phase.

\subsection{UASD}
\vspace{-5pt}
We denote the embeddings extracted by the audio encoder as $\z = E\left( \x \right)$,
where $E(\cdot)$ is the audio encoder.
Then, the anomaly score of a test sample $\x^\text{te}$ is given as the knn distance between the test sample and the training data measured in the embedding space, which is
\begin{align}
    \mathcal{A} \left( \x^{\text{te}} \right)
    &=
    \frac{1}{k}
    \sum_{i=1}^{k} d
    \left(
        E\left(\x^\text{te}\right),
        E\left(\x^\text{tr}_{n_i}\right)
    \right),
    \label{eq: anom score}
\end{align}
where $d(\cdot,\cdot)$ is a function that measures the distances between the given embeddings, such as the Euclidean distance or the Cosine distance.
$n_i~(i=1,\dots,k)$ is the index of the training data of which $\x_{n_i}^\text{tr}$ is the $i$-th nearest neighbor training sample in terms of $d(\cdot, \cdot)$.
This strategy of combining an audio encoder with a knn-based anomaly score calculator has been used in various UASD methods~\cite{Kawaguchi2021description, dohi2022description}, including methods used in top rankings of the latest UASD competitions~\cite{LvAITHU2024, JiangTHUEE2024}.
This means that the proposed method can be used to extend such methods for timbre difference capturing, which is a further advantage of the proposed method.

\subsection{Timbre difference capturing}
Let $T_l\left(\x\right)~(l=1, \dots, L)$ be the timbre metric value of attribute $l$ for an audio sample $\x$, which is computed by the timbral models.
We estimate the timbre difference label by evaluating how much the timbre metric of the test sample deviates from the knn normal training samples.
Suppose $T_l\left(\x^\text{te}\right)$ was the $r$-th smallest value among
$\{T_l\left(\x^\text{te}\right), T_l\left(\x_{n_1}^\text{tr}\right), \cdots, T_l\left(\x_{n_k}^\text{tr} \right)\}$.
Then, we compute the timbre difference score of $\x^\text{te}$ as
\begin{align}
    \hat{y}_l \left( \x^\text{te} \right) = \frac{r-1}{k} \in \left[0, 1\right]. \label{eq: label score}
\end{align}
This evaluates how large the timbre metric of the test sample is among the $k$ nearest neighbors training samples in a nonparametric manner.
Note that this value is equivalent to the special case of the U value used in the Mann-Whitney U test~\cite{mason2002areas}, with normalization.
The Mann-Whitney U test is a nonparametric statistical test that evaluates whether the given two sets of samples are sampled from different distributions.
Therefore, this value can be used for evaluating differences in the timbre metric values.
Lastly, by using a predefined threshold $t\in [0,1]$, the timbre difference labels are estimated as
\begin{align}
    \hat{y}_{l} \left( \x^\text{te} \right) =
    \begin{cases}
        -1 & \hat{y}_{l} \leq t \\
        0 & t < \hat{y}_{l} < 1-t\\
        1 & 1-t \leq \hat{y}_{l}.
    \end{cases}
    \label{eq: estimate y}
\end{align}

\begin{table}[!t]
    \centering
    \caption{Statistics of ground truth timbre difference labels. \#g, \#u, r denote number of conditions, number of unique timbre difference label vector, and number of labels of each values.}
    \scriptsize
    \vspace{-6pt}
    \scalebox{0.85}{
    \begin{tabular}{@{}c@{~}|ccc|ccc|ccc@{}}
        \hline
        Section  & \multicolumn{3}{c}{section 00} & \multicolumn{3}{c}{section 01}         & \multicolumn{3}{c}{section 02} \\
        Machine  & \# g & \# u & r ($-1/0/1$)& \# g & \# u & r ($-1/0/1$)& \# g & \# u & r ($-1/0/1$)\\
        \hline
        Bearing  & 26 & 23 & 32/49/34 & 32 & 19 & 16/44/35  & 4 & 1 & 2/3/0 \\
        Fan      & 3 & 2 & 1/6/3 & 7 & 6 & 7/6/17 & 3 & 3 & 5/9/1 \\
        Gearbox  & 26 & 17 & 14/28/43 & 23 & 13 & 6/15/44 & 6 & 6 & 8/9/13 \\
        Slider   & 26 & 11 & 24/14/17 & 26 & 10 & 26/15/9 & 6 & 4 & 4/11/5 \\
        Valve    & 4 & 2 & 2/7/1 & 8 & 8 & 3/29/8 & 7 & 4 & 9/2/9 \\
        \hline
    \end{tabular}
    }
    \label{tab:ground truth}
    \vspace{-15pt}
\end{table}

\section{Dataset creation}
\vspace{-5pt}

For evaluating the proposed method, we created a UASD dataset with ground truth timbre difference labels.
For the audio data, we used the MIMII DG dataset~\cite{Dohi2022mimiidg}, which is a UASD dataset with domain generalization problem settings included.
The whole or a subset of this dataset have been used in recent UASD competitions, the DCASE Challenge Task 2 in 2022 -- 2024~\cite{dohi2022description, Dohi2023description, nishida2024description}, for the development dataset.
There are five machine types with three sections of data for each machine type, each section including source and target domain data.
While the MIMII DG dataset consists of machine sounds corrupted with factory noise, we also own the originally recorded machine sounds and additional information on the cause of anomalies for each anomalous sample.
We used these original data and information to assign the ground truth labels. 
For clarity, we will refer to the originally recorded machine sound data with no factory noise as the "clean dataset`` and the MIMII DG dataset as the "noisy dataset``.
For further details of the dataset, please refer to \cite{Dohi2022mimiidg}.

The ground truth timbre difference labels for anomalous sounds were automatically generated using the timbre metrics computed by timbral models.
To only extract the sound difference specifically caused by machine anomaly, the ground truth labels should be determined by comparing the anomalous sounds with normal sounds that are recorded under identical conditions, including both machine operational and recording conditions.
Furthermore, if the cause of the anomaly is identical, the ground truth label should also be identical.
Therefore, we initially determined a single ground truth label for each condition and cause of anomalies, and then assigned these same label values to anomalous samples with identical conditions and causes.

For a single data section in the clean dataset, suppose there are $M$ conditions and $Q$ types of anomaly causes.
Here, the conditions can be the machine's operational conditions such as machine speed, recording conditions such as microphone locations, or a combination of them.
Let $\mathcal{D}_m^\text{tr}$ and $\mathcal{D}_{m, q}^{\text{anom}}, m=1,\cdots, M, q=1,\dots,Q$, be a subset of  normal training data in the clean dataset that was recorded under condition $m$ and a subset of the anomalous data in the clean dataset for condition $m$ and anomaly cause $q$, respectively.
To derive the ground truth label for condition $m$ of anomaly cause $q$,
we compute a score that indicates how much each timbre attribute differ between these two sets of audio samples.
Let $\mathcal{T}_l\left(\D \right)=\{T_l(\x) | \x\in \D\}$ denote the values of timbre metrics of timbre attribute $l$ for a set of audio samples $\{\D\}$.
We then evaluate the deviation of the timbre metrics between normal and anomalous samples as
\begin{align}
    \tilde{y}_{m,q} = \text{AUC}\left( \mathcal{T}_l\left(\D_m^\text{tr} \right), \mathcal{T}_l\left(\D_{m,q}^\text{anom} \right) \right) \in [0,1].
\end{align}
Here, $\text{AUC}\left( \mathcal{T}_1, \mathcal{T}_2 \right)$ denotes the area under the receiver operating characteristic curve~(AUC) when $\mathcal{T}_1$ and $\mathcal{T}_2$ are regarded as the scores of negative and positive samples, respectively.
Note that AUCs are also equivalent to the normalized version of the U value in the Mann-Whitney U test ~\cite{mason2002areas}, which justifies using this score.
Next, by conducting the same thresholding shown in \eqref{eq: estimate y} by another predefined threshold $t'\in [0,1]$, we obtain the ground truth timbre difference label $y_{m,q}\in\{1,0,-1\}$.
After observing the scores of $\tilde{y}_{m,q}$, we set $t'=0.05$.
Finally, we assign each anomalous sample in the noisy dataset~(=MIMII DG dataset) the labels computed for the corresponding condition and anomaly cause.
Note that in inference, we cannot run the same procedure that created these ground truth labels, since conditions of the test data are unknown and the data is corrupted with environmental noise.
This is why we have to estimate the timbre difference label such as in the proposed method.

We summarize how many unique label combinations were created and the ratio of each label value in TABLE~\ref{tab:ground truth}.
The variety of the ground truth labels indicates that estimating these labels can be informative in explaining the difference between normal and anomalous sounds.
For example, one damage type in Bearing section 00 resulted in increased Sharpness and Boominess for a specific velocity, whereas another damage type resulted in decreased Sharpness and Brightness for the same velocity.
Thus, by interviewing machine inspectors beforehand, it might be possible to automatically distinguish the anomaly causes using the acquired information.
\setlength{\tabcolsep}{1mm} 
\begin{table}[t]
\begin{center}
\caption{AUC for each method (\%). Due to space constraints, we only show mean AUC for each machine in target domain.}
\label{tab:auc}
\scriptsize
\scalebox{0.85}{
\begin{tabular}{@{}cc@{~}l@{~}| c c c c c@{}}
\hline
\\[-6.5pt]
Domain & Machine & Sec. &
Timbre-knn &
Mbnv2 &
PANNs &
CLAP  &
BEATs\\[-2.5pt]

& & & &
(Proposed) &
(Proposed) &
(Proposed) &
(Proposed)
\\[-1.5pt]
\hline
&         & 00 & 68.0 & 62.5 &  74.1 & 62.0 & \bf{78.2}\\
& Bearing & 01 & 62.9 & \bf{79.9} & 59.7 & 65.7 & 62.6\\
&         & 02  & 57.7 & \bf{69.8} & 59.8 & 68.4 & 63.7\\
    \cline{2-8}
&         & 00  & 47.3 & 62.9 & 75.7 & 67.7 & \bf{86.2}\\
&     Fan & 01  & 68.2 & \bf{74.5} & 68.2 & 71.3 & 69.4\\
&         & 02  & 77.7 & 68.2 & 74.3 & 70.1 & \bf{80.9}\\    
     \cline{2-8}
&         & 00  & 52.8 & 69.6 & 45.8 & 69.9 & \bf{80.2}\\
source & Gearbox & 01  & \bf{56.5} & 56.1 & 53.6 & \bf{56.5} & 52.5\\
&         & 02  & 57.3 & 82.5 & 74.9 & 73.2 & \bf{86.6}\\
         \cline{2-8}
&         & 00  & 75.6 & 81.0 & 76.0 & \bf{85.8} & 85.6\\
&  Slider & 01  & 55.7 & 61.9 & 70.3 & \bf{85.9} & 80.5\\
&         & 02  & 73.1 & 64.2 & 86.5 & \bf{89.7} & 77.6\\
         \cline{2-8}
&         & 00  & 58.1 & \bf{90.4} & 52.6 & 63.3 & 57.0\\
&   Valve & 01  & 47.2 & \bf{59.4} & 50.3 & 53.3 & 51.0\\
&         & 02  & 83.7 & \bf{86.8} & 70.0 & 64.0 & 57.5\\
         \cline{2-8}
&  \multicolumn{2}{c|}{mean} & 62.8 & \bf{71.3} & 66.1 & 69.8 & \bf{71.3}\\
  \hline \hline
& \multicolumn{2}{c|}{Bearing} & 49.7 & \bf{63.7} & 55.4& 52.8 & 48.4 \\
& \multicolumn{2}{c|}{Fan} & 45.1 & 47.7 & 48.2 & 45.4 & \bf{54.8}\\
target & \multicolumn{2}{c|}{Gearbox} & 52.5 & \bf{68.8} & 56.7 & 61.3 & 65.8\\
& \multicolumn{2}{c|}{Slider} & 54.3 & 56.2 & 58.5 & \bf{70.0} & 60.5\\
& \multicolumn{2}{c|}{Valve} & 55.6 & 57.7 & \bf{59.5} & 58.2 & 50.0\\
\cline{2-8}
& \multicolumn{2}{c|}{mean} & 51.4 & \bf{58.8} & 55.7 & 57.5 & 55.9 \\
    \hline
\end{tabular}
}
\vspace{-25pt}
\end{center}
\end{table}
\begin{figure*}[!t]
	\begin{center}
        \includegraphics[width=0.8\hsize,clip]{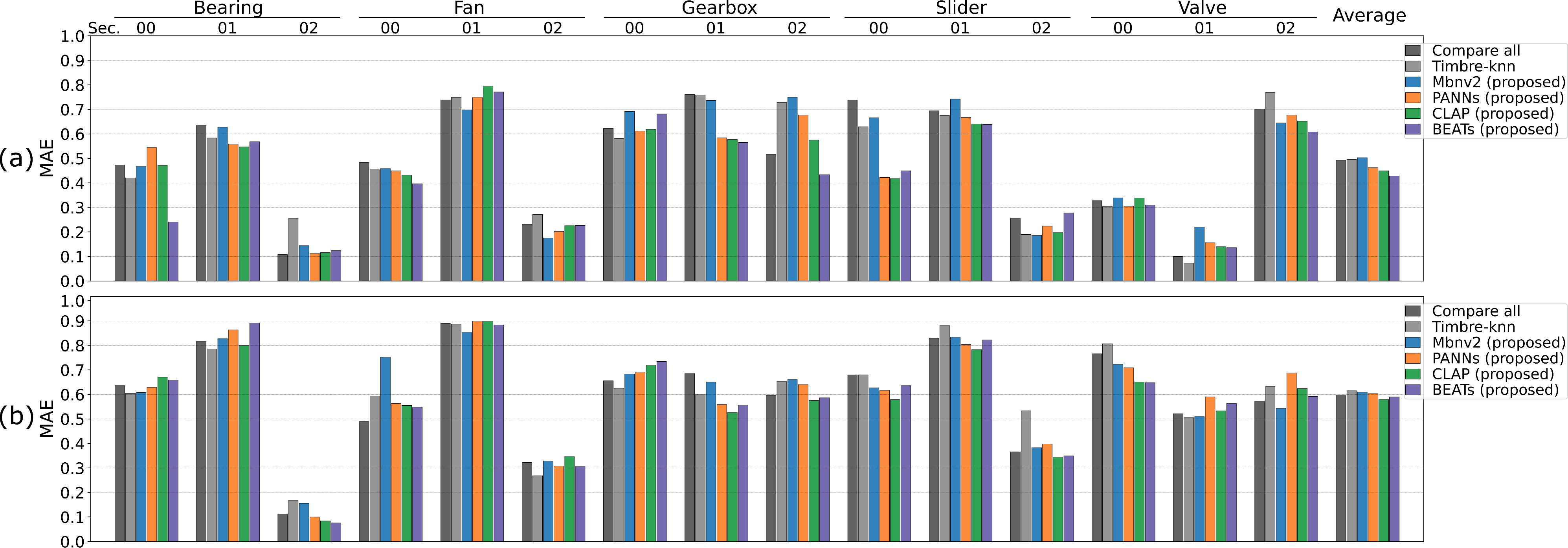}
	\caption{MAE of timbre difference capturing~(smaller is better). (a) Source domain, (b) Target domain}
	\label{fig:mae}
        \vspace{-18pt}
	\end{center}
\end{figure*}

\section{Experiments}
\vspace{-7pt}
\subsection{Experimental conditions}
\vspace{-5pt}
To evaluate the proposed method's effectiveness, we conducted a joint UASD and timbre difference capturing experiment on the dataset described earlier.
For the audio encoder, we employed four models: 
MobileNet-v2~\cite{Sandler2018}, PANNs~\cite{kong2020panns}, audio encoder of CLAP~\cite{Elizalde2024clap}, and BEATs~\cite{Chen2023BEATs}.
MobileNet-v2 was the baseline model in DCASE 2022~\cite{dohi2022description} and BEATs~\cite{Chen2023BEATs} featured in some of the top-ranking solution for the DCASE 2024 Challenge task 2~\cite{LvAITHU2024, JiangTHUEE2024}.
For MobileNet-v2, the same preprocessing as in the baseline model~\cite{dohi2022description} was conducted for the input data and it was trained by a classification task for machine attributes given in the MIMII DG dataset for each machine type.
The model was optimized with AdamW with a learning rate of $10^{-4}$ for $50$ epochs.
For the other models, we used the publicly available pretrained model~\cite{kong2020panns, Elizalde2024clap, Chen2023BEATs} without modification.
For all models, we set $k=30$. 
However, similar results were observed when $k$ was set around $10$ to $40$.

We also conducted two methods as baselines.
In the first, we estimated the timbre difference labels by comparing the anomalous sample's timbre metrics with all normal training samples using \eqref{eq: label score}.
This served as the baseline for timbre difference label estimation without the proposed method.
In the second method, we conducted UASD and timbre difference capturing using the $L$ timbre metrics as feature vectors for knn, instead of the audio embeddings~(Timbre-knn).
Comparing with this method allowed us to evaluate the effectiveness of using audio encoder embeddings.

The UASD performance was evaluated by AUC.
The timbre difference capturing performance was evaluated by the mean absolute error~(MAE) for anomalous samples, where we normalized the error to compensate for imbalances in ground truth labels~\cite{baccianella2009evaluation}.
That is,
\begin{align}
	\mbox{MAE}_{l} = \frac{1}{3}\sum_{i=1}^{M} \frac{|\hat{y}_{l} \left( \x^\text{te}_{i} \right) - y_{l, i}|}{M_{l, y_{l,i}}},
\end{align}
for timbre $l$, where $\{\x_i^\text{te}\}_{i=1}^M$ are the anomalous test data, $y_{l,i}$ is the ground truth timbre difference label for timbre $l$ of sample $\x_i$, $M_{l, y}=|\{ y_{l,i} \mid y_{l,i}=y, 1\leq i \leq M \}|$ is the number of samples in the test data that the ground truth label value is identical to $y$.

\subsection{Results}

For reference, we show the AUC values of each method in TABLE~\ref{tab:auc}. 
All four audio encoders had higher average AUC values than Timbre-knn. 
This means that audio embedding spaces from these models can better distinguish between normal and anomalous samples than the predefined timbre metric space, aligning with the proposed method's concept shown in Fig.~\ref{fig:method_abs}.

Next, we present the MAE of timbre difference capturing in Fig.~\ref{fig:mae}. 
In the source domain, PANNs, CLAP, and BEATs showed smaller average MAE compared to the baseline methods, showing the effectiveness of the proposed method. 
Notably, BEATs had significantly smaller MAE on several machines and sections, resulting in the smallest overall MAE.
Most of the those sections, such as Bearing 00, Gearbox 01, and Slider 00, were those with a high variety of conditions~(see TABLE~\ref{tab:ground truth}).
This indicates that the original motivation of the proposed method to accurately estimate timbre differences even when normal sounds have a high variety, has been accomplished to some extent.
BEATs also had generally high AUC values, suggesting that a good embedding space can accomplish such motivation.
In contrast, MobileNet-v2 had the same average AUC value as BEATs but a larger average MAE than the baseline methods. 
This might be because MobileNet-v2 was trained only on normal target machine sounds and consequently lacked the ability to correctly measure the similarity between them and out-of-distribution anomalous data, only having the ability to distinguish whether a given sample in out-of-distribution or not. 
Conversely, other audio encoders were trained on a wide variety of sound data, potentially resulting in embedding spaces that better capture similarity between normal and anomalous machine sounds.
In the target domain, there was little difference in MAE, possibly due to the small data size. 
This is out of this paper's scope and measures will be necessary in the future.
Overall, by using suitable audio encoders that are presumed to have good embedding spaces, the proposed method can accurately estimate timbre differences even when normal data have a variety, provided there is enough training data.

\section{Conclusion}
\vspace{-5pt}
We proposed a framework to explain anomaly differences based on predefined timbre attributes, which does not require anomalous data or ground truth labels for training.
We then further proposed a joint UASD and timbre difference capturing method that compares anomalous test samples only with the most similar normal training samples.
We achieved this by using audio encoders and k-nearest neighbors in the embedding space the encoders generate.
Experiments using the MIMII DG-based dataset, which includes ground truth timbre difference labels, confirmed
that the proposed method can accurately estimate the timbre difference labels even when the normal training sounds have a variety due to various conditions.

\vfill\pagebreak
\bibliographystyle{IEEEtran}
\bibliography{str_def_abrv,refs}

\end{document}